\documentclass[,final]
  {aipproc}

\layoutstyle{6x9}
\begin{document}

\title{Diffraction Slopes for Elastic Proton-Proton and Proton-Antiproton Scattering}


\author{V.A. Okorokov}{
  address={Moscow Engineering Physics Institute (State University), 115409,
Moscow, Russia} }

\begin{abstract}
The diffraction slope parameter is investigated for elastic
proton-proton and proton-antiproton scattering based on the all
available experimental data at low momentum transfer values.
Energy dependence of the elastic diffraction slopes is
approximated by various analytic functions. The expanded
"standard" logarithmic approximations allow to describe
experimental slopes in all available energy range reasonably.
Various approximations differ from each other both in the low
energy and very high energy domains. Predictions for diffraction
slope parameter are obtained for elastic proton-proton scattering
at RHIC and LHC energies, for proton-antiproton elastic reaction
in FAIR energy domain for various approximation functions.
\end{abstract}

\maketitle

\section{Introduction}

Elastic hadron-hadron scattering, the simplest type of hadronic
collision process, remains one of the topical theoretical problems
in fundamental interaction physics at present. In the case of $pp$
and $\bar{p}p$ elastic scattering, although many experiments have
been made over an extended range of initial energies and momentum
transfer, these reactions are still not well understood. The
nuclear slope $B$ for elastic scattering is of interest in its own
right. This quantity defined according to the following equation:
\begin{equation}
B\left(s\right)=\left.\left[\frac{\textstyle
\partial}{\textstyle
\partial t}\left(\ln \frac{\textstyle \partial \sigma\left(s,t\right)}{\textstyle
\partial t}\right)\right]\right|_{t=0}, \label{Slope-def}
\end{equation}
is determined experimentally. On the other hand the study of $B$
parameter is important, in particular, for  reconstruction
procedure of full set of helicity amplitudes for elastic nucleon
scattering \cite{Okorokov-arXiv-0711.2231}. The present status of
slope for elastic $pp$ and $\bar{p}p$ scattering is discussed over
the full energy domain.

\section{Slope energy dependence}

We have attempted to describe the energy behaviour of the elastic
nuclear slopes for $pp$ and $\bar{p}p$ reactions. Specifically, we
have focused on the low $|t|$ domain. The ensemble of experimental
data on slopes for elastic nucleon collisions can be fitted
successfully by many phenomenological models, at least for
$\sqrt{s} > 5$ GeV. The new "expanded" logarithmic
parameterizations have been suggested in
\cite{Okorokov-arXiv-0711.2231} for description of the elastic
slope at all available energies. Thus taking into account standard
Regge parameterization and quadratic function of logarithm from
\cite{Block-RevModPhys-57-563-1985} the following analytic
equations are under study here: \vspace*{-0.25cm}
$$
\begin{array}{llr}
\hspace*{3.0cm} B\left(s\right)=&
B_{0}+2a_{1}\ln\left(s/s_{0}\right),&~~~~~~~~~~~~~~~~~~~~~~~~~
(2a)
\label{Fit-1} \\
\hspace*{3.0cm} B\left(s\right)=&
B_{0}+2a_{1}\ln\left(s/s_{0}\right)+a_{2}\left
[\ln\left(s/s_{0}\right)\right]^{a_{3}}, & (2b)
\label{Fit-2}\\
\hspace*{3.0cm} B\left(s\right)=&
B_{0}+2a_{1}\ln\left(s/s_{0}\right)+a_{2}\left(s/s_{0}\right)^{a_{3}},
& (2c) \label{Fit-3}\\
\hspace*{3.0cm} B\left(s\right)=&
B_{0}+2a_{1}\ln\left(s/s_{0}\right)+a_{2}\left
[\ln\left(s/s_{0}\right)\right]^{2}, & (2d) \label{Fit-4}
\end{array}\vspace*{-0.25cm}
$$
where $s_{0}=1$ GeV$^{2}$, in general case parameters $B_{0},~
a_{i}, i=1,2,3$ depend on range of $|t|$ which is used for
approximation.

Experimental data are from \cite{Lasinski-NPB-37-1-1972,
Ambast-PRD-9-1179-1974, Jenni-NPB-129-232-1977,
Fajardo-PRD-24-46-1981, Iwasaki-NPA-433-580-1985,DDG-url}. Total
number of experimental points is equal 133 / 129 for $pp /
\bar{p}p$ scattering at low $|t|$, respectively. Thus the
experimental sample is significantly larger than that for some
early investigations \cite{Okorokov-arXiv-0711.2231,
Block-RevModPhys-57-563-1985}. As known the measurements of
nuclear slope, especially at intermediate $|t|$ do not form a
smooth set in energy, in contrast to the situation for global
scattering parameters $\rho$ and $\sigma_{tot}$, where there is a
good agreement between various group data
\cite{Block-RevModPhys-57-563-1985}. Thus the data samples for
approximations are somewhat smaller because of exclusion of points
which differ significantly from the other experimental points at
close energies. The maximum fraction of excluded points is equal
3.1\% for low $|t|$ domain.

The energy dependence for experimental slopes and corresponding
fits by (2a)--(2d) are shown at Fig.1a and Fig.1b for $pp$ and
$\bar{p}p$ correspondingly. The fitting parameter values are
indicated in Table 1. The available systematic errors were added
in quadrature to the statistical errors at calculation of
resulting errors for fitted data points. One can see that the
fitting functions (2a), (2d) describe the $pp$ (Fig.1a) and
$\bar{p}p$ (Fig.1b) experimental data statistically acceptable
only for $\sqrt{s} \geq 5$ GeV. The parameter $a_{1}$ for function
(2a) agree with estimate for Pomeron intercept $\alpha'_{\cal{P}}
\approx 0.25$ GeV$^{-2}$ for $\bar{p}p$ data but this parameter is
some larger than $\alpha'_{\cal{P}}$ for proton-proton data. One
can see the $a_{2}$ parameter is equal zero within errors for
approximation of $\bar{p}p$ data by (2d). Thus the function (2d)
can be excluded from the set of approximation functions for
$\bar{p}p$ at low $|t|$ values. The RHIC point for proton-proton
collisions has a large error and can't discriminates the
approximations. The parameterizations (2b), (2c) allow to describe
experimental data at all energies with reasonable fit quality for
$pp$. The functions (2a)--(2c) are close to each other at
$\sqrt{s} \geq 5$ GeV, especially, the Regge model approximation
and (2c) fit. It seems the ultra-high energy domain is suitable
for separation of various parameterizations. The qualities of
(2b), (2c) approximations for $\bar{p}p$ elastic scattering data
are much poorer because of very sharp behaviour of experimental
data near the low energy boundary. But one can see that the
functions (2b), (2c) agree with experimental points at qualitative
level and (very) close to each other for all energy range.
\begin{figure}
\begin{tabular}{cc}
\mbox{\includegraphics[width=7.5cm,height=7.0cm]{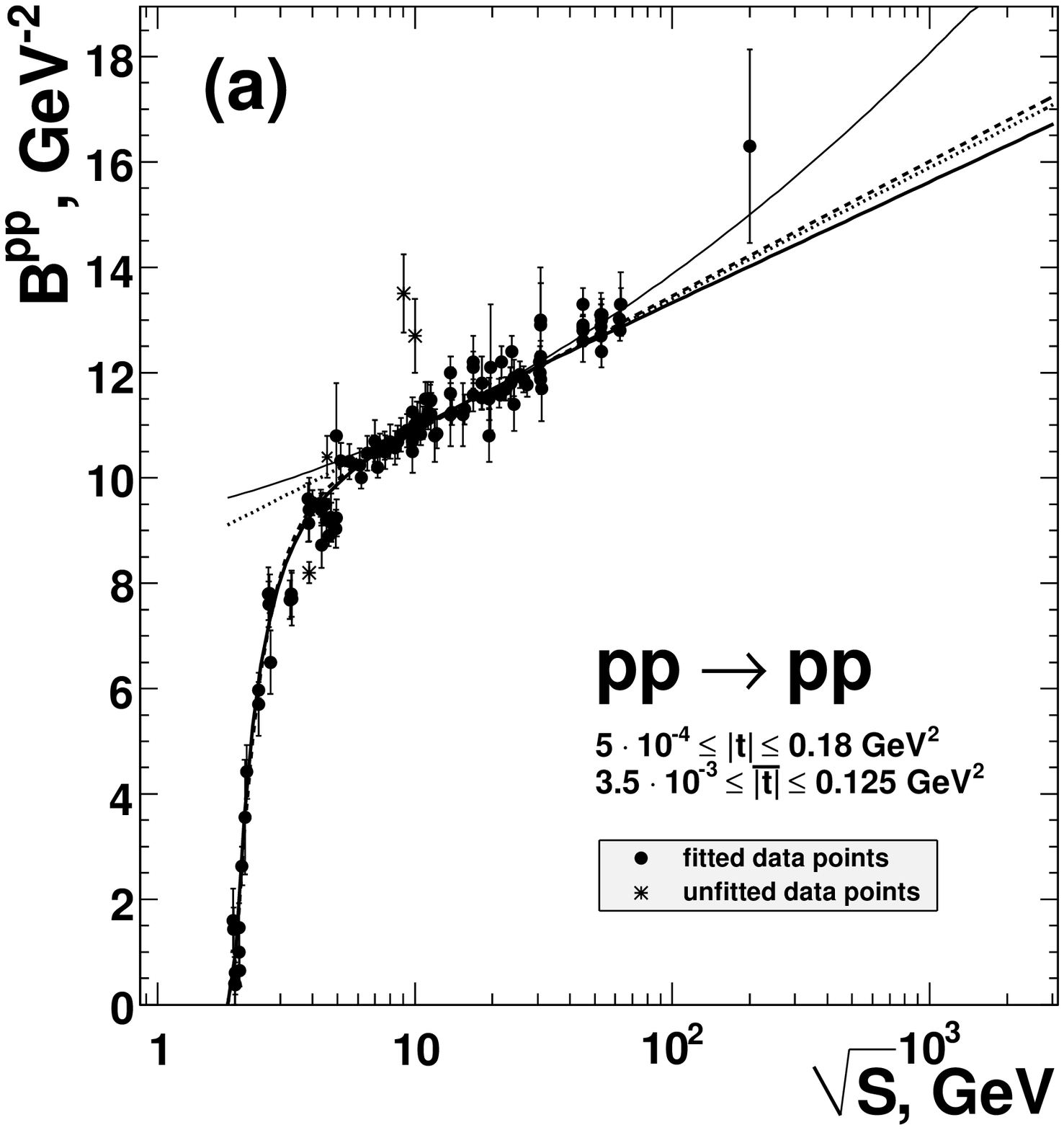}}&
\mbox{\includegraphics[width=7.5cm,height=7.0cm]{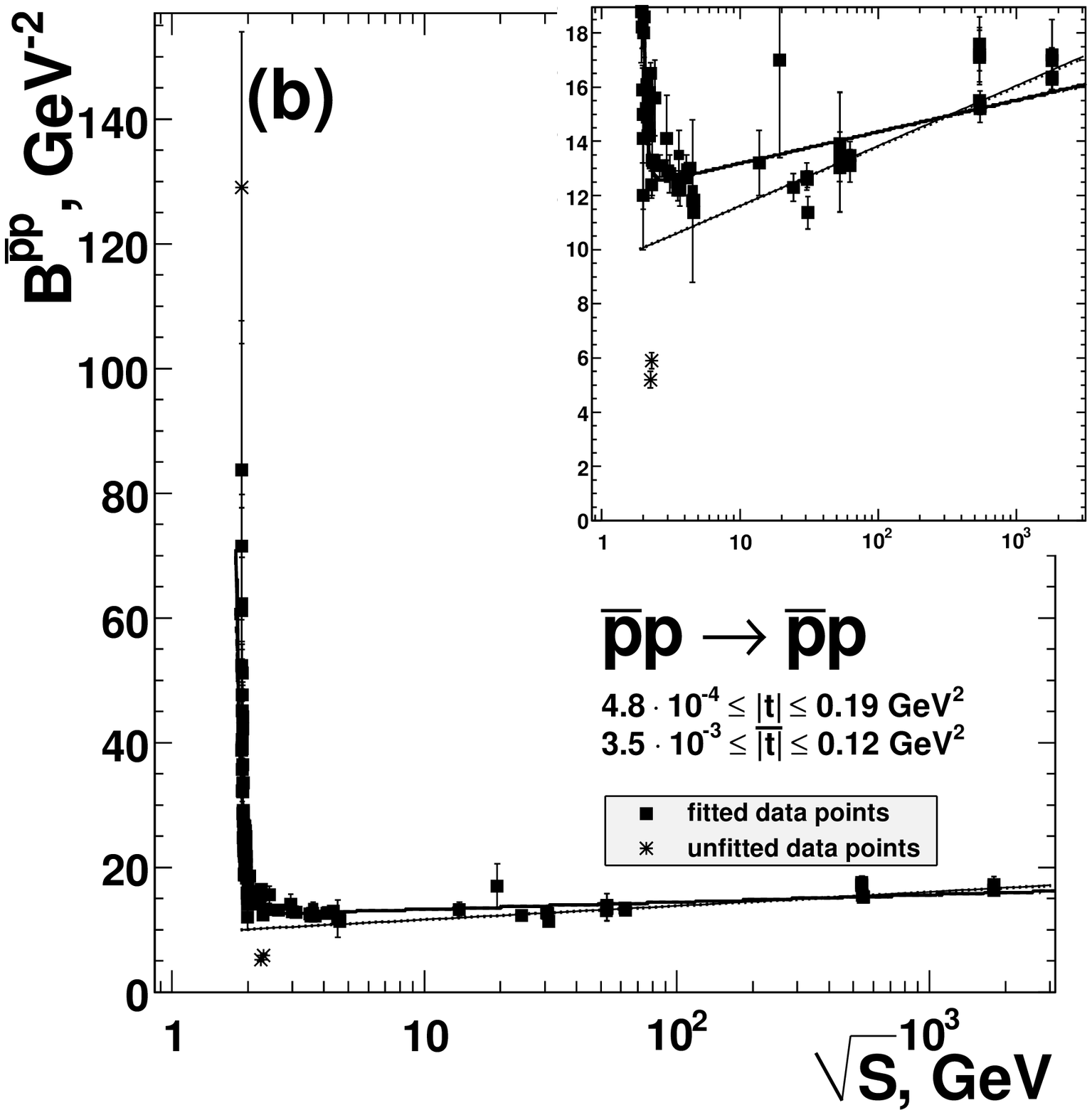}}
\end{tabular}
\caption{Energy dependence of the elastic slope parameters for
proton-proton (a) and proton-antiproton (b) scattering for low
$|t|$ domain. The inner picture for (b) shows the experimental
data and fits at the same scale as well as for (a). The curves
correspond to the fitting functions as following: (2a) -- dot,
(2b) -- thick solid, (2c) -- dot-dashed, (2d) -- thin.}
\end{figure}
\begin{table}
\begin{tabular}{lccccc}
\hline \multicolumn{1}{l}{Function} &
\multicolumn{5}{c}{Parameter} \\
\cline{2-6} \rule{0pt}{10pt}
 & $B_{0}$, GeV$^{-2}$ & $a_{1}$, GeV$^{-2}$ & $a_{2}$, GeV$^{-2}$ & $a_{3}$ & $\chi^{2}/\mbox{n.d.f.}$ \\
\hline
\multicolumn{6}{c}{proton-proton scattering} \\
\hline
(2a) & $8.43 \pm 0.08$  & $0.270 \pm 0.007$ & --              & --               & $162/87$ \\
(2b) & $8.77 \pm 0.12$  & $0.248 \pm 0.009$ & $-27 \pm 2$     & $-3.45 \pm 0.14$ & $329/125$ \\
(2c) & $8.33 \pm 0.08$  & $0.278 \pm 0.007$ & $-181 \pm 23$   & $-2.21 \pm 0.09$ & $340/125$ \\
(2d) & $9.3 \pm 0.3$    & $0.11 \pm 0.06$   & $0.03 \pm 0.01$ & --               & $155/86$ \\
\hline
\multicolumn{6}{c}{proton-antiproton scattering} \\
\hline
(2a) & $9.4 \pm 0.4$    & $0.24 \pm 0.02$   & --              & --               & $29/22$ \\
(2b) & $11.98 \pm 0.06$ & $0.127 \pm 0.006$ & $489 \pm 97$    & $-13.0 \pm 0.5$  & $1180/122$ \\
(2c) & $12.01 \pm 0.06$ & $0.125 \pm 0.005$ & $\left(2.7 \pm 1.4\right) \cdot 10^{6}$& $-9.3 \pm 0.4$ & $1260/122$ \\
(2d) & $9.4 \pm 1.6$    & $0.24 \pm 0.16$   & $\left(2 \pm 157\right) \cdot 10^{-4}$ & -- & $29/21$ \\
\hline
\end{tabular}
\caption{Fitting parameters for slope energy dependence in low
$|t|$ domain} \label{tab:1}
\end{table}

One can get a predictions for nuclear slope parameter values for
some facilities based on the results shown above. The $B$ values
at low $|t|$ for different energies of FAIR, RHIC, and LHC are
shown in the Table 2. According to the results above the function
(2d) are not considered for $\bar{p}p$ collisions.
\begin{table}
\begin{tabular}{lccccccccc}
\hline \multicolumn{1}{l}{Fitting} &
\multicolumn{9}{c}{Facility energies, $\sqrt{s}$} \\
\cline{2-10} \rule{0pt}{10pt} function & \multicolumn{4}{c}{FAIR,
GeV} &
\multicolumn{2}{c}{RHIC, TeV} & \multicolumn{3}{c}{LHC, TeV} \\
\cline{2-10} \rule{0pt}{10pt}
 & 3 & 5 & 6.5 & 14.7 & 0.2 & 0.5 & 14 & 28 & 42\tablenote{The ultimate energy upgrade of LHC project \cite{Skrinsky-ICHEP2006}} \\
\hline
(2a) & -- & 10.95 & 11.20 & 11.98 & 14.15 & 15.14 & 18.74 &
19.49 & 19.93 \\
(2b) & 12.56 & 12.80 & 12.93 & 13.35 & 14.02 & 14.93 & 18.24 &
18.93 & 19.33 \\
(2c) & 12.56 & 12.81 & 12.95 & 13.35 & 14.22 & 15.24 & 18.95 &
19.72 & 20.17 \\
(2d) & -- & -- & -- & -- & 15.00 & 16.67 & 24.44 & 26.39 & 27.58 \\
\hline
\end{tabular}
\caption{Predictions for nuclear slope based on the
parameterizations (2a) -- (2d) for low $|t|$ domain} \label{tab:2}
\end{table}
As expected the functions (2b) and (2c) predicted the very close
slope parameter values for FAIR. The first approximation function
(2a) can predicts for $\sqrt{s} \geq 5$ GeV only. Functions
(2a)--(2c) predict much smaller values for $B$ in high-energy $pp$
collisions than (2d) approximation. Perhaps, the future more
precise RHIC results will agree better with predictions based on
experimental data fits under study. It should be emphasized that
the fits (2a)--(2c) of available experimental data predict the
slower increasing of $B$ with energy than most of phenomenological
models \cite{Kundrat-EDS-273-2007}. The $B$ values predicted for
LHC at $\sqrt{s}=14$ TeV by (2a) and (2c) are most close to the
some model expectations \cite{Block-PRD-60-054024-1999,
Petrov-EPJ-C28-525-2003}. Moreover one needs to emphasize that the
model estimates at $\sqrt{s}=14$ TeV were obtained for
$B\left(t=0\right)$ and the $t$-dependence of slope shows the
slight decreasing of $B$ at growth of momentum transfer up to $t
\approx 0.1-0.2$ GeV$^{2}$ in particular for the models
\cite{Block-PRD-60-054024-1999, Petrov-EPJ-C28-525-2003}. Thus one
can expect the some better agreement between model estimates and
predicted values of $B$ from Table 2 for finite (non-zero) low
$|t|$ values. Possibly the saturation regime, Black Dick Limit,
will be reached at the LHC. The one of the models in which such
effects appear, namely, DDM predicts the slope $B\left(t=0\right)
\approx 23$ GeV$^{-2}$ at $\sqrt{s}=14$ TeV
\cite{Selyugin-EDS2007-279}.

\section{Summary}

The main results of this paper are the following. Most of all
available experimental data for slope parameter in elastic nucleon
collisions are approximated by different analytic functions. The
suggested parameterizations allow to describe experimental nuclear
slope at all available energies in low $|t|$ domain for $pp$ quite
reasonably. The new approximations agree with experimental
$\bar{p}p$ data at qualitative level but these fits are still
statistically unacceptable because of very sharp behavior of $B$
near the low energy limit. Predictions for slope parameter are
obtained for elastic proton-proton and proton-antiproton
scattering in energy domains of some facilities.

\end{document}